\documentclass[aps,superscriptaddress,twocolumn,showpacs,preprintnumbers,amsmath,amssymb,newcommand]{revtex4}
\usepackage[cp1251]{inputenc}
\usepackage{amssymb,amsmath}
\usepackage[mathscr]{eucal}
\usepackage{graphicx}
\usepackage{longtable}
\usepackage[dvips]{hyperref} 
\begin{document}

\title{Quantum-mechanical Landau-Lifshitz equation}

\author{Y.Yerchak (a), D.Yearchuck (b) \\
\textit{(a) - Belarusian State University, Nezavisimosti Ave.4, Minsk, 220030, RB; Jarchak@gmail.com,\\ (b) - Minsk State Higher College, Uborevich Str.77, Minsk, 220096, RB; dpy@tut.by}}

\date{\today}
             
\begin{abstract}
Quantum-mechanical analogue of Landau-Lifshitz equation has been derived. It has been established that Landau-Lifshitz equation is fundamental physical equation underlying the dynamics of spectroscopic transitions and transitional phenomena. New phenomenon is predicted: electrical spin wave resonance (ESWR) being to be electrical analogue of magnetic spin wave resonance. 
\end{abstract}

\maketitle
Correct formal description of dynamics of spectroscopic transitions as well 
as a number of transitional effects, that is, for instance, 
Rabi-oscillations, free induction and spin echo effects in magnetic 
resonance spectroscopy and their optical analogues in optical spectroscopy 
is achieved in the frame of a gyroscopic model, see, e.g., \cite{Macomber, Scully}. 
Mathematical base for gyroscopic model is Landau-Lifshitz (L-L) 
equation. L-L equation was postulated by Landau and Lifshitz for macroscopic 
classical description of the motion of magnetization vector in ferromagnets 
as early as 1935 \cite{Landau_Lifshitz_1935}. L-L equation is as follows:
\begin{equation}
\label{eq1}
\frac{d\vec {S}}{dt}=[\gamma _{_H} \vec {H}\times \vec {S}],
\end{equation}
where $\vec {S}$ is magnetic moment, $\vec {H}$ is effective magnetic field, 
$\gamma _{_H} $ is gyromagnetic ratio. L-L equation was substantiated 
quantum-mechanically in magnetic resonance theory, but as the equation, 
describing the only the motion of magnetic moment in external magnetic 
field. At the same time L-L equation and Bloch equations, which are based on 
L-L equations, were in fact postulated for description of \textit{dynamics} of magnetic 
resonance \textit{transitions}, see, e.g., \cite{Slichter}. The gyroscopic model for optical 
transition dynamics and for description of optical transitional effects was 
introduced formally on the base of analogy with gyroscopic model, developed 
for magnetic resonance. However, the optical analogue of L-L equation was 
obtained quantum-statistically by means of density operator formalism \cite{Macomber, Scully}:
\begin{equation}
\label{eq2}
\frac{d\vec {P}}{dt}=[\vec {P}\times \gamma _{_E} \vec {E}],
\end{equation}
where $\vec {P}$, $\vec {E}$ are vectors, which are considered to 
be some mathematical abstractions, since their components represent themselves various physical quantities. So, in \cite{Macomber} is emphasized that the only $P_{x}$, 
$P_{y}$ and $E_{x}$, $E_{y}$ components of vectors $\vec {P}$, $\vec {E}$, correspondingly, 
characterize the genuine electromagnetic properties of the system, at the same time the components $P_{z}$ and $E_{z}$ cannot be reffered to electromagnetic characteristics. Further,
$\gamma _{_E}$ was called gyroelectric ratio the only tentatively, its analogy with gyromagnetic ratio was 
 suggested. The formal character of optical gyroscopic model is indicated also in modern quantum optics theory, see, e.g., 
\cite{Scully}. This situation is consequence of the attempt to preserve the only polar symmetry properties for the vectors of electric field strength and electric polarization of the medium. At the same time the vectors $\vec {P}$, $\vec {E}$ in eq.2 should be axial vectors. To built the axial vectors, satysfying eq.2, the third components of polar vectors of electric field strength and electric polarization is changed artificially into quantities which are believed to be some mathematical abstractions in the sence, that they cannot be reffered to electric field strength and electric polarization correspondingly. So, it is generally accepted, that the components of the vectors $\vec {P}$, $\vec {E}$ in eq.2 are consisting from various physical quantities. This situation seems to be incorrect, if to proceed from the assumption of the identity of the nature of the spectroscopic transitions in optical and radiospectroscopy regions.

The aim of given work is to obtain quantum-mechanical equations for 
description of dynamics of both magnetic resonance and optical transitions with clear physical sence of all the quantities 
(for the case of simple 1D model of quantum system). 

Let us consider the general properties of electromagnetic 
field to clarify the sense of the quantities in (\ref{eq2}). It is well known, that 
electromagnetic field can be characterized by both contravariant tensor 
$F^{\mu \nu }$ (or covariant $F_{\mu \nu })$ and contravariant pseudotensor 
$\tilde {F}^{\mu \nu }$ which is dual to $F^{\mu \nu }$ (or 
covariant $\tilde {F}_{\mu \nu }$, which is dual to $F_{\mu \nu })$. For 
instance, $\tilde {F}^{\mu \nu }$ is determined by the following relation: $\tilde {F}^{\mu \nu }=\frac{1}{2}e^{\alpha \beta \mu \nu }F_{\mu \nu }$, where $e^{\alpha \beta \mu \nu }$ is Levi-Chivita fully antisymmetric unit 
4-tensor. The use of field tensors and pseudotensors seems to be equally possible 
by description of electromagnetic field and its interaction with a matter. 
So, for example, by using of both field tensor and pseudotensor the 
field invariants are obtained \cite{Landau_Lifshitz_Field_Theory, Stephani}. Further in the practice of treatment of experimental results is generally accepted, that electric field strength, dipole 
moment and polarization vectors are polar vectors but magnetic 
field strength, dipole moment and magnetization vectors are always 
axial vectors. However for the effects, describing the interaction of electromagnetic field in optical experiments 
the picture seems to be reverse. It follows directly from the structure of algebraic linear space, which produce field of 
tensors and pseudotensors. Really, if the structure of $F^{\mu \nu }$ is\begin{equation}
\label{eq3}
F^{\mu \nu }=\left[ {{\begin{array}{rrrr}
 0  & {-E_1 }  & {-E_2} & {-E_3} \\
 {E_1 }  & {0}  & {-H_3} & {H_2} \\
 {E_2 }  & {H_3} & {0}  & {-H_1} \\
 {E_3 }  & {-H_2} & {H_1} & {0}  \\
\end{array} }} \right]
\end{equation}
then the structure of $\tilde {F}_{\mu \nu }$ will be:
\begin{equation}
\label{eq4}
\tilde {F}^{\mu \nu }=\left[ {{\begin{array}{rrrr}
 0 & {-H_1 } & {-H_2 } & {-H_3 }\\
 {H_1 } & {0} & {E_3 } & {-E_2 }\\
 {H_2 } & {-E_3 } & {0} & {E_1 }\\
 {H_3 } & {E_2 } & {-E_1 } & {0}\\
\end{array} }} \right].
\end{equation}
So we have for contravariant and covariant electromagnetic field 
pseudotensors the expressions: 
\begin{equation}
\label{eq5}
\tilde {F}^{\mu \nu }=(-\vec {H},\vec {E}),
\quad
\tilde {F}_{\mu \nu } =(\vec {H},\vec {E}).
\end{equation}
We consider further formally the same vectors $\vec {H}$ and $\vec {E}$, that is, the vectors consisting of the components, determined by the structure of $\tilde {F}^{\mu \nu}$, but now vector $\vec {H}$ is polar and vector $\vec {E}$ is axial. The possibility of given consideration requires the validation, which can be obtained from the following. 
Let us define the space F of sets of contravariant tensors $\left\{{F}_{\mu \nu }\right\}$ and pseudotensors $\left\{\tilde {F}_{\mu \nu }\right\}$ and corresponding to them sets of covariant tensors and pseudotensors of electromagnetic field over the field of scalar values P. It is evident that all the axioms of linear space hold true, i.e., if ${F_1}^{\mu \nu}$ and ${F_2}^{\mu \nu}$ $\in F$, 
then 
\begin{equation}\ {F_1}^{\mu \nu} + {F_2}^{\mu \nu} = {F_3}^{\mu \nu} \in F, \end{equation}
and, if 
$F^{\mu \nu }\in F$,
then
\begin{equation}\ \alpha\ F^{\mu \nu } \in F \end{equation} for $\forall \alpha\ \in P$. 
Let us define the linear algebra $\mathfrak F$ by means of definition in above definited vector space $\left\langle F,+,\cdot \right\rangle$ of transfer operation $(\ast)$ to dual tensor, using the convolution with Levi-Chivita fully antisymmetric unit 
4-tensor $e_{\alpha \beta \mu \nu }$. It is evident, that in algebra $\left\langle \mathfrak F,+,\cdot, \ast \right\rangle$ the following axioms of linear algebra hold true:
if ${F_1}^{\alpha \beta}$ and ${F_2}^{\alpha \beta}$ $\in \mathfrak F$, 
then 
\begin{equation} e_{\mu \nu\alpha \beta} ({F_1}^{\alpha \beta} + {F_2}^{\alpha \beta}) = \tilde {(F_1)}_{\mu \nu } + \tilde {(F_2)}_{\mu \nu } \in \mathfrak F, \end{equation}
\begin{equation} ({F_1}^{\alpha \beta} + {F_2}^{\alpha \beta})e_{\alpha \beta \mu \nu }= {F_1}^{\alpha \beta} e_{\alpha \beta \mu \nu }+ {F_2}^{\alpha \beta} e_{\alpha \beta \mu \nu } \in \mathfrak F,\end{equation}
\begin{equation} (e_{\alpha \beta \mu \nu } \lambda {F}^{\mu \nu}) = \lambda (e_{\alpha \beta \mu \nu } {F}^{\mu \nu}) = (e_{\alpha \beta \mu \nu } {F}^{\mu \nu}) \lambda \end{equation} for $\forall \lambda\ \in P$. We can determine now on the space $\left\langle F,+,\cdot \right\rangle$ the functional $\Phi$ as follows: \begin{equation} \Phi({F}^{\mu \nu}) \equiv \left\langle {F}^{\mu \nu} | \Phi\right\rangle = {F}^{\mu \nu} \tilde {F}_{\mu \nu }, \end{equation} and 
\begin{equation} \Phi (\tilde {F}^{\mu \nu}) \equiv \left\langle \tilde {F}^{\mu \nu} | \Phi \right\rangle = \tilde {F}^{\mu \nu} {F}_{\mu \nu }. \end{equation} In other words, the mapping $\Phi:{{F}^{\mu \nu}, \tilde {F}^{\mu \nu}} \rightarrow \tilde {P}$ is taking place, where $\tilde {P}$ is pseudoscalar field. It is clear, that $\Phi$ on the space F is linear functional. Really, let $\alpha, \beta \in P$, then, taking into account the properties (8) to (10) of $\left\langle \mathfrak F,+,\cdot, \ast \right\rangle$, we have 
\begin{equation} \left\langle {F}^{\mu \nu} | \alpha \Phi_1 + \beta \Phi_2 \right\rangle = \alpha \left\langle {F}^{\mu \nu} | \Phi_1 \right\rangle + \left\langle {F}^{\mu \nu} | \Phi_2 \right\rangle, \end{equation} \begin{equation} \left\langle \tilde {F}^{\mu \nu} | \alpha \Phi_1 + \beta \Phi_2 \right\rangle = \alpha \left\langle \tilde {F}^{\mu \nu} | \Phi_1 \right\rangle + \left\langle \tilde {F}^{\mu \nu} | \Phi_2 \right\rangle. \end{equation} Consequently, the set ${\Phi({F}^{\mu \nu}, \tilde {F}^{\mu \nu})}$ of linear functionals on the space $\left\langle F,+,\cdot \right\rangle$ represents itself linear space over the field P (since it is evident, that the conditions like to (6) and (7) hold true), which is dual to space $\left\langle F,+,\cdot \right\rangle$. Therefore, we have \begin{equation} \left\langle \Phi,+,\cdot \right\rangle = \left\langle {F}^{\times},+,\cdot \right\rangle. \end{equation} It is substantionally, that ${F}^{\times}$ is not self-dual. Actually, the "vector" (in algebraic sense), which can be built on basis "vectors" of space F with the projections, which are functionals, corresponding, in accordance with relationships (11), (12), to given basis "vectors" of F, cannot belong to F. Really, its components are pseudoscalars and, being to be considered as coefficients in linear combination of the elements of vector space F, cannot belong to field P (properties (6) (7) do not hold true). At the same time full physical description of dynamical system requires to take into consideration both the spaces (more strictly, Gelfand three should be considered, if corresponding topology is determined in these spaces). On other hand, the description with help of Gelfand three will be equivalent to the following description. We define starting space F over the P + $\tilde {P}$. Then resulting functional space will be self-dual. In extended by such a way space we can choose 4 physically different subspaces: two subspaces 1) $\left\{{F}^{\mu \nu }\right\}$ and 2) $\left\{\tilde {F}^{\mu \nu }\right\}$ over the scalar field P and two subspaces 3) $\left\{{F}^{\mu \nu }\right\}$ and 4) $\left\{\tilde {F}^{\mu \nu }\right\}$ over the pseudoscalar field $\tilde {P}$. The second case differs from the first case by the following. Symmetry properties of $\vec {E}$ and $\vec {H}$ remain the same, i.e., $\vec {E}$ is polar vector, since it is dual vector to antisymmetric 3D pseudotensor, and $\vec {H}$, respectively, is axial. At the same time, the components of vector $\vec {E}$ correspond now to pure space components of field tensor ${\tilde {F}^{\mu \nu}}$, the components of vector $\vec {H}$ correspond to time-space mixed components. Arbitrary element of the third subspace \begin{equation} \alpha {F}^{\mu \nu }(x_1) + \beta {F}^{\mu \nu }(x_2), \end{equation} where $\alpha, \beta \in \tilde {P}, x_1, x_2$ are the points of Minkowski space, 
 represents itself the 4-pseudotensor. Its space components, being to be the components of antisymmetric 3-pseudotensor, determine dual polar vector $\vec {H}$, mixed components are the components of 3-pseudovector $\vec {E}$. Therefore, the symmetry properties of the components of vectors $\vec {E}$ and $\vec {H}$ relatively the improper rotations will be inverse to the case 1. It is evident, that in the 4-th case the symmetry properties of the components of vectors $\vec {E}$ and $\vec {H}$ relatively the improper rotations will be inverse to the case 2.
 Given consideration allows to suggest, that free electromagnetic field is 4-fold degenerated. The interaction with device (or, generally, with some substance) relieves degeneracy and leads to formation of resonance state: field + device (substance), which has finite lifetime. It is reasonable to suggest, that the field in the resonance state becomes nondegenerate. Realization of concrete field state (one of 4 possible) will, evidently, be determined by symmetry characteristics of registering device (interacting substance). So, we suggest, that, in principle, various symmetry properties of the same field can be obtained by registration of interaction with the same substance, but with various methods, e.g., with ESR and optical absorption. 
 
We will consider the case 4 in more detail.
The components of $\vec {H}$ and $\vec {E}$ are determined in this case
 by other potentials, accordingly, by dual scalar $\tilde {\varphi }$ 
and dual vector $\tilde {\vec {A}}$ potentials, which can be obtained from the solution of the 
following set of differential equations:
\begin{equation}
\label{eq6}
\begin{split}
\nabla \tilde {\varphi }(\vec {r},t)=&-[\nabla \times \vec {A}\,(\vec 
{r},t)], \\
[\nabla \times \tilde {\vec {A}}\,(\vec {r},t)]=&\frac{\partial \vec 
{A}\,(\vec {r},t)}{\partial x^0}-\frac{\partial A_{\,0} (\vec 
{r},t)}{\partial \vec {r}},
\end{split}
\end{equation}
where $A_{0} (\vec {r},t) \equiv \varphi (\vec {r},t)$ is scalar 
potential and $\vec {A}\,(\vec {r},t)$ is vector potential, which determine 
the components of genuine field tensors $F^{\mu \nu }$, F$_{\mu \nu }$. It 
can be shown, that the solution of eq.\ref{eq6} is viewed as follows:
\begin{equation}
\label{eq7}
\tilde {\varphi }=-\int\limits_o^{\vec {r}} {[\nabla \times \vec 
{A}]} \,d\vec {r}, \
\tilde {\vec {A}}=\tilde {\vec {A}}_1 +\tilde {\vec 
{A}}_{\,2},
\end{equation}
where 
\begin{equation}
\label{eq8}
\tilde {\vec {A}}_1 =-\frac{1}{4\pi}\nabla \left({\int\limits_{(\vec {\rho })} {\frac{Q}{\vert \vec {r}-\vec {\rho }\vert}} \,d^3\rho } \right)\Rightarrow \tilde {\vec {A}}_1 =\frac{Q}{4\pi }\frac{\vec 
{r}}{\vert \vec {r}\vert ^2},
\end{equation}
\begin{equation}
\label{eq9}
\tilde {\vec {A}}_{2} =
\left[ {\nabla \times \frac{1}{4\pi }\int\limits_{(\vec {\rho })} 
{\frac{\frac{\partial \vec {A}}{\partial x^0}-\frac{\partial 
A_0 }{\partial \vec {\rho }}}{\vert \vec {r}-\vec {\rho 
}\vert }} \,d^3\rho } \right].
\end{equation}
It is taken into account by derivation of expressions (\ref{eq8}) and (\ref{eq9}), that 
dual vector potential satisfies to calibration condition: $\nabla \cdot 
\tilde {\vec {A}}(\vec {r},t)=Q$, where $Q$ is const. It is evident, that in the case of usually used Coulomb calibration dual vector potential will be determined the only by relationship (\ref{eq9}).
Thus the simple analysis shows, that, if electric field components are 
components of \textit{pseudotensor}, the equation of dynamics of \textit{optical} transitions will have 
mathematically the same structure with that one for the equation of 
dynamics of \textit{magnetic resonance} transitions (in which magnetic field components are parts of
genuine tensor $F^{\mu \nu }$). In other words, mathematical abstractions in 
eq.\ref{eq2} become real physical sense: $\vec{E}$ is vortex part of intracrystalline and 
external electric field, $\vec{P}$ is electrical moment, which should be defined like 
to magnetic moment. So, both the vectors are axial vectors, that is, they 
have necessary symmetry properties (relatively reflection and inversion), in 
order to satisfy the eq.\ref{eq2}. It should be noted that the statement on "equality 
in rights" of genuine field tensor and pseudotensor by 
description of electromagnetic phenomena follows from general consideration 
of the \textit{geometry} of Minkowski space, which determine unambiguously the full set of 3 possible kinds of 
geometrical objects. Really, the tensors and pseudotensors are equally 
possible geometrical objects among them for 
any pseudo-Euclidean abstract space, to which Minkowski space is 
isomorphic \cite{Rashevskii}.
 
Quantum-mechanical description of dynamics of spectroscopic transitions on the example of 1D 
system interacting with electromagnetic field confirm given general 
conclusion. Really, let us consider the system representing 
itself the periodical ferroelectrically (ferromagnetically) ordered chain of 
$n$ equivalent elements, interacting with external oscillating electromagnetic 
field. It is assumed that the interaction between elements (elementary 
units) of the chain can be described by the Hamiltonian of quantum XYZ 
Heisenberg model in the case of a chain of magnetic dipoles and by 
corresponding optical analogue of given Heisenberg model in the case of a 
chain of electric dipoles. We will consider for the simplicity the case of 
isotropic exchange. Each elementary unit of the chain will be considered as 
two-level system like to one-electron atom. Then Hamiltonians for the 
chain of electrical dipole moments and for the chain of magnetic dipole 
moments will be mathematically equivalent. We will use further the rotating wave approximation \cite {Scully}. Then the chain for distinctness of 
electrical dipole moments can be described by the following Hamiltonian:
\begin{equation}
\label{eq10}
\begin{split}
\raisetag{40pt}
&\mathcal{\hat H} = \frac{\hbar \omega _0}{2}\sum\limits_n {\hat {\sigma}_n^z } - p_{_{E}}^{\alpha \beta }\sum\limits_n {E_1^n (\hat {\sigma} _n^ + e^{ - i\omega t} + \hat {\sigma} _n^- e^{i\omega t} )}\\
& \ \  + \sum\limits_n [J_{_{E}} (\hat {\sigma} _n^ + \hat {\sigma} _{n + 1}^- + \hat {\sigma} _n^ -  \hat {\sigma} _{n + 1}^ + + \frac{1}{2}\hat {\sigma} _n^z \hat {\sigma} _{n + 1}^z ) + H.c.].
\end{split}
\end{equation}
where $\hat {\sigma }_n^z =\left| {\alpha _n } \right\rangle \left\langle 
{\alpha _n } \right|-\left| {\beta _n } \right\rangle \left\langle {\beta _n 
} \right|$ is so called $\hat {\sigma }_z $-operator, observable quantity 
for which is population difference of the states of $n$-\textit{th} element, $\hat 
{\sigma }_n^+ =\left| {\alpha _n } \right\rangle \left\langle {\beta _n } 
\right|$ and $\hat {\sigma }_n^- =\left| {\beta _n } \right\rangle 
\left\langle {\alpha _n } \right|$ are transition operators of $n$-\textit{th} element 
from eigenstate $\left| {\alpha _n } \right\rangle $ to eigenstate $\left| 
{\beta _n } \right\rangle $ and vice versa, correspondingly. It is suggested 
in the model, that $\left| {\alpha _n } \right\rangle $ and $\left| {\beta _n } \right\rangle $ are eigenstates, producing the full set for each of $n$ elements. It is evident, that given assumption can be realized strictly the 
only by the absence of the interaction between the elements. At the same 
time proposed model will rather well describe the real case, if the 
interaction energy of adjacent elements is much less of the energy of the 
splitting $\hbar \omega _0 =\mathcal{E}_\beta -\mathcal{E}_\alpha$ between the energy levels, 
corresponding to the states $\left|\alpha_n\right\rangle$ and $\left|\beta_n\right\rangle$. This case includes in fact all known 
experimental situations. Further, $p_{_E}^{\alpha \beta }$ is matrix element 
of dipole transitions between the states $\left| {\alpha _n } \right\rangle $ and $\left| {\beta _n } \right\rangle $, which along with energy difference of these states $\mathcal{E}_\beta -\mathcal{E}_\alpha$ are suggested to be 
independent on $n$, $E_1^n $ is amplitude of electric component of 
electromagnetic wave on the $n$\textit{-th} element site, $\hbar$ is Planck's constant, $J_{E}$ is optical analogue of the 
exchange interaction constant. Here, in correspondence with the suggestion, 
$J_{_E} =J_{_E}^x =J_{_E}^y =J_{_E}^z $. In the case of the 
chain of magnetic dipole moments $E_1^n $ in Hamiltonian (\ref{eq10}) is replaced by $H_1^n 
$, i.e., by amplitude of magnetic component of electromagnetic wave on the 
$n$\textit{-th} element site, $J_{_E}$ is replaced by the exchange interaction constant 
$J_{_H}$, matrix element $p_{_E}^{\alpha \beta }$ is replaced by $p_{_H}^{\alpha \beta} $ and the 
frequency $\omega_0 $ is replaced by $ \frac{1}{\hbar}g_{_H} \beta_{_H} H_0 = \gamma_{_H} H_0 $, where $H_0$ is external static magnetic field, $\beta_{_H}$ is Bohr magneton, $g_{_H}$ is $g$-tensor, which is assumed for the simplicity to be isotropic. The first term in Hamiltonian (\ref{eq10}) characterizes the total energy of all chain elements in the absence of external field and in the absence of interaction 
between chain elements. The second item characterizes an interaction of a 
chain with an external oscillating electromagnetic field in dipole 
approximation. Matrix elements of dipole transitions $p_{_E}^{\alpha \beta } $ 
and $p_{_E}^{\beta \alpha } $ between couples of the states ($\left|{\alpha_n} \right\rangle $, $\left|{\beta_n} \right\rangle$) and ($\left|{\beta_n} \right\rangle$, $\left|{\alpha_n} \right\rangle $), respectively, are assumed to be equal, i.e., spontaneous emission is not taken into consideration. The third item is, in essence, Hamiltonian of quantum Heisenberg \textit{XXX}-model in the case of magnetic 
version and its electrical analogue in electric version of the model 
proposed. Let us define the vector operator: 
\begin{equation}
\label{eq11}
\hat {\vec {\sigma }}_k =\hat {\sigma }_k^- \,\vec {e}_+ +\hat {\sigma 
}_k^+ \,\vec {e}_- +\hat {\sigma }_k^z \,\vec {e}_z .
\end{equation}
It seems to be the most substantial for the subsequent analysis, that a set 
of $\hat {\sigma }_k^m $ operators, where m is z, +, -, produces algebra, 
which is isomorphic to $S = 1/2$ Pauli matrix algebra, i.e., 
mappings $f_k :\,\hat {\sigma }_k^m \to \sigma _P^m $ realize 
isomorphism. Here $k$ is a number of chain unit, $\sigma _P^m $ 
is the set of Pauli matrices for the spin of $1/2$. Consequently, from 
physical point of view $\hat {\vec {\sigma }}_k$ represents itself some 
vector operator, which is proportional to operator of the spin of $k$-\textit{th} chain 
unit. Vector operators $\hat {\sigma }_k^m $ produce also linear space over complex field, which 
can be called transition space. It is 3-dimensional (in the case of two-level systems), that is 3 operator 
equations of the motion for components of $\hat {\vec {\sigma }}_k$ is necessary for correct 
description of optical transitions. Typical inaccuracy in many of the quantum 
optics calculations of transition dynamics is connected with obliteration of correct dimensionality of transition space.
The equation of the motion for $\hat {\vec {\sigma }}_k$ is:
\begin{equation}
\label{eq12}
i\hbar \,\frac{{\partial \hat {\vec {\sigma} }_k}}{{\partial t}} = [\hat {\sigma} _k^ -  ,\mathcal{\hat H} ]\, \vec e_ +   + [\hat {\sigma} _k^ +  ,\mathcal{\hat H} ]\, \vec e_ -   + [\hat {\sigma} _k^z ,\mathcal{\hat H} ]\, \vec e_z. 
\end{equation}
We will consider the case of homogeneous excitation of the chain, that is $E_1^n $ is independent on unit number $n$ ($E_1^n \equiv E_1)$. Then, using the commutation relations
$[\hat {\sigma }_k^+ ,\hat {\sigma }_n^- ]=\delta _{kn} \hat {\sigma }_n^z 
,\,[\hat {\sigma }_n^- ,\hat {\sigma }_k^z ]=2\delta _{kn} \hat {\sigma 
}_n^- ,\,[\hat {\sigma }_k^z ,\hat {\sigma }_n^+ ]=2\delta _{kn} \hat 
{\sigma }_n^+$,
we obtain
\begin{subequations}
\label{eq13}
\begin{gather}
\frac{{\partial \hat\sigma _k^z }}{{\partial t}} = 2i\Omega _{_E} \left( {e^{ - i\omega t} \hat\sigma _k^ + - e^{i\omega t} \hat\sigma _k^ - } \right) 
\\ 
+\frac{{2iJ_{_E} }}{\hbar }\left(\left\{ {\hat\sigma _k^ - } \right.,\left. {(\hat\sigma _{k + 1}^ + + \hat\sigma _{k - 1}^ + )} \right\} - 
\left\{ {\hat\sigma _k^ + } \right.,\left. {(\hat\sigma _{k + 1}^ - + \hat\sigma _{k - 1}^ - )} \right\}\right),\nonumber
\\
\frac{{\partial \hat\sigma _k^ +  }}{{\partial t}} = i\omega _0 \hat\sigma _k^ +   + i\Omega _{_E}  e^{i\omega t} \hat\sigma _k^z 
\\
+\frac{{iJ_{_E} }}{\hbar } \left(\left\{ {\hat\sigma _k^ +  } \right.,\left. {(\hat\sigma _{k + 1}^z  + \hat\sigma _{k - 1}^z )} \right\} - \left\{ {\hat\sigma _k^z } \right.,\left. {(\hat\sigma _{k + 1}^ +   + \hat\sigma _{k - 1}^ +  )} \right\}\right),\nonumber
\\
\frac{{\partial \hat\sigma _k^-}}{{\partial t}} =  - i\omega _0 \hat\sigma _k^ -   - i\Omega _{_E} e^{ - i\omega t} \hat\sigma _k^z  
\\  
-\frac{{iJ_{_E} }}{\hbar }   \left(\left\{ {\hat\sigma _k^ -  } \right.,\left. {(\hat\sigma _{k + 1}^z  + \hat\sigma _{k - 1}^z )} \right\} + \left\{ {\hat\sigma _k^z } \right.,\left. {(\hat\sigma _{k + 1}^ - + \hat\sigma _{k - 1}^ - )} \right\}\right),\nonumber
\end{gather}
\end{subequations}
where expressions in braces $\left\{ {\,,\,} \right\}$ are anticommutants. Here 
$\Omega _{E}$ and $\gamma _{_E} $ is Rabi frequency and gyroelectric ratio. 
They are determined, correspondingly, by relations 
$\Omega _{_E} =\frac{E_1 p_{_E}^{\alpha \beta } }{\hbar }=\gamma _{_E} E_1 ,
\quad
\gamma _{_E} =\frac{p_{_E}^{\alpha \beta } }{\hbar }$, which are
 replaced by relations $\Omega _{_H} =\frac{H_1 p_{_H}^{\alpha \beta }}{\hbar}=\gamma_{_H} H_1 $,
$\gamma_{_H}=\frac{p_{_H}^{\alpha \beta }}{\hbar}$ in the case 
of the chain of magnetic dipole moments. The equations (\ref{eq13}) can be represented in compact vector 
form, at that the most simple expression is obtained by using of the basis 
$\vec {e}_+ =(\vec {e}_x +i\vec {e}_y ), \,\vec {e}_- =(\vec {e}_x -i\vec{e}_y ), \vec {e}_z $. So, we have
\begin{equation}
\label{eq14}
\frac{{\partial \hat {\vec {\sigma}} _k }}{{\partial t}} = \left[ {\hat {\vec {\sigma}} _k \times \hat {\vec {\mathfrak{G}}}_{k - 1,k + 1} } \right],
\end{equation}
where $\hat {\vec {\sigma }}_k$ is given by (\ref{eq11}), but with the components, 
corresponding to new basis, $k=\overline {2,N-1} $, and vector operator 
$\hat {\vec {\mathfrak{G}}}_{k - 1,k + 1}  = \hat {\mathfrak{G}}_{k - 1,k + 1}^-  \vec e_ +   + \hat {\mathfrak{G}}_{k - 1,k + 1}^ +  \vec e_ -   + \hat {\mathfrak{G}}_{k - 1,k + 1}^z \vec e_z$, where its components are
\begin{subequations}
\label{eq15}
\begin{gather}
\hat {\mathfrak{G}}_{k - 1,k + 1}^-  = \Omega _{_E} e^{-i\omega t} - \frac{{2J_{_E} }}{\hbar }(\hat\sigma _{k + 1}^- + \hat\sigma _{k - 1}^-),  \\
\hat {\mathfrak{G}}_{k - 1,k + 1}^+ = \Omega _{_E} e^{i\omega t} - \frac{{2J_{_E} }}{\hbar }(\hat\sigma _{k + 1}^+ + \hat\sigma _{k - 1}^ + ),  \\
\hat {\mathfrak{G}}_{k - 1,k + 1}^z = - \omega_0 - \frac{{2J_{_E} }}{\hbar }(\hat\sigma _{k + 1}^z + \hat\sigma _{k - 1}^z ).
\end{gather}
\end{subequations}
It should be noted, that vector product of vector operators in (\ref{eq14}) is 
calculated in the correspondence with known expression
\begin{equation}
\label{eq16}
\left[ {\hat {\vec {\sigma}} _k \times \hat {\vec {\mathfrak{G}}}_{k - 1,k + 1} } \right] = \left| {\begin{array}{*{20}c}
 {\vec e_- \times \vec e_z} & {\hat{\sigma}_k^-} & {\hat {\mathfrak{G}}_{\,\,k - 1,k + 1}^-} \\
 {\vec e_z \times \vec e_+} & {\hat{\sigma}_k^+} & {\hat {\mathfrak{G}}_{\,\,k - 1,k + 1}^+} \\
 {\vec e_+ \times \vec e_-} & {\hat{\sigma}_k^z} & {\hat {\mathfrak{G}}_{\,\,k - 1,k + 1}^z} \\
\end{array}} \right|,
\end{equation}
however, by its calculation one should take anticommutants of corresponding 
components instead their products. 
Given definition seems to be natural generalization of vector product for 
the case of operator vectors, since the only in this case the result will 
 be independent on a sequence of components of both the vectors in their 
products like to that one for usual vectors. Naturally, the expressions like 
to (\ref{eq16}) can be used for calculation of vector product of common 
vectors. Taking into account the physical sense of vector operators $\hat 
{\vec {\sigma }}_k$ we conclude, that (\ref{eq14}) represent themselves 
required quantum-mechanical difference-differential equations (the time is 
varied continuously, the coordinates are varied discretely) for the 
description of the dynamics of the spectroscopic transitions (in the frames 
of the model proposed). From here in view of isomorphism of algebras of 
operators $\hat {\vec {\sigma }}_k $ and components of the spin it follows 
that the (\ref{eq14}) is equivalent to L-L equation (in its 
difference-differential form). Consequently we have proved the possibility 
to use L-L equation for the description of the dynamics of spectroscopic 
transitions, as well as for the description of transitional effects. To 
obtain the continuous approximation of (\ref{eq14}) for coordinate variables too, 
we have to suggest that the length of electromagnetic wave $\lambda $ 
satisfies the relation: $\lambda >> a$, where $a$ is 1D-chain-lattice constant. Then the 
continuous limit is realized if to substitute all the operators, which 
depend on discrete variable $k$, for the operators depending on continuous 
variable $z$, that is: $\hat{\sigma }_k^\pm \to \hat{\sigma }^\pm 
(z), \hat{\sigma }_k^z \to \hat{\sigma }^z(z)$. Thus, we obtain, 
taking also into account the relations $\hat{\sigma }_{k+1}^{z,\pm } 
+\hat{\sigma }_{k-1}^{z,\pm } -2\hat{\sigma }_k^{z,\pm } \to 
a^2\frac{\partial ^2\hat\sigma^{z,\pm } (z)}{\partial z^2}$, the 
equation, which, like to (\ref{eq14}), in compact vector form is:
\begin{equation}
\label{eq17}
\frac{\partial \hat {\vec {\sigma }}(z)}{\partial t}=\left[ {\hat {\vec 
{\sigma }}(z)\times \gamma _{_E} \vec {E}} \right]-\frac{2a^2J_{_E} }{\hbar 
}\left[ {\hat {\vec {\sigma }}(z)\times \nabla ^2\hat {\vec {\sigma }}(z)} 
\right],
\end{equation}
where $\vec {E}=E_{1} e^{i\omega t}\vec {e}_- +E_{1} e^{-i\omega t}\vec{e}_+ +\left( {\frac{-\omega _0 }{\gamma _{_E} }} \right)\vec {e}_z $.
The structure of vector $\vec {E}$ clarifies its physical meaning. Two components 
$E^+,\,\,E^-$ are right- and left-rotatory components of oscillating external 
electric field, third component $E^z$ is intracrystalline electric field, 
which produces two level energy splitting for each of the unit of a chain 
system with value, equal to $\hbar \omega _0 $. It means that the following 
relation is taking place: $\omega _0 =\frac{1}{\hbar }g_{_E} \beta _{_E} E_0 
=\gamma _{_E} E_0 $, where $\beta _{_E} $ is electrical analogue of Bohr magneton, 
$g_{_E} $ is electrical analogue of magnetic $g$ -tensor, which is assumed for the 
simplicity to be isotropic. In other words, by means of given relation the 
correspondence between an unknown intracrystalline electric field $E_0 $ and 
observed frequency $\omega _0 $ is set up.
Further, we take into consideration, that physical sense of operators $\hat 
{\vec {\sigma }}(z)$ in continuous limit is remained, i.e., for each point 
of $z$ the components $\widehat{\sigma }^\pm (z)$, $\widehat{\sigma }^z(z)$ are 
satisfying to algebra, which is isomorphic to algebra of the set of spin \
components. Then by means of relation, which has mathematically the same 
form for both the types of the systems studied $\hat {\vec {S}}(z)\sim \frac{\hbar \hat {\vec {\sigma }}(z)}{2}$, the equation of the motion for operators of magnetic and electric spin moments are obtained. So, e.g., the equation of the motion for electric spin moment operator is:
\begin{equation}
\label{eq18}
\frac{\partial \hat {\vec {S}}(z)}{\partial t}=\left[ {\hat {\vec 
{S}}(z)\times \gamma _{_E} \vec {E}} \right]-\frac{4a^2J_{_E}}{\hbar ^2}\left[ {\hat 
{\vec {S}}(z)\times \nabla ^2\hat {\vec {S}}(z)} \right]
\end{equation}
Equation (\ref{eq18}) gives for the case $J = 0$ quantum-mechanical optical analogue 
of classical Landau-Lifshitz equation in continuous limit, in fact it is 
operator equation, which argues the correctness of eq.(\ref{eq2}), that is, the 
physical correctness of gyroscopic model for description of optical 
transitions and transitional optical analogues of magnetic resonance 
phenomena. If $J \ne 0$ we have quantum-mechanical optical analogue of 
classical L-L equation, which was introduced by Kittel for SWR description 
\cite{Kittel}. Therefore, the results obtained allow to predict a new phenomenon - 
electric spin wave resonance (ESWR). The equation (\ref{eq18}) (if put aside the 
operator symbols) and equation introduced by Kittel are coinciding 
mathematically to factor 2 in the second term. This difference is like to 
well known difference of gyromagnetic ratios for orbital and spin angular 
moments. 
Thus along with magnetic resonance methods we can detect a spin value of 
particles, quasiparticles, impurities or other centers in solids by optical 
methods: by study of transitional optical analogues of magnetic resonance 
phenomena or ESWR. It should also be noted that above considered theoretical 
description of ESWR allow predict the difference in splitting constants 
which characterize ESWR by its experimental detection with using of 
one-photon methods like to IR-absorption or IR-reflection and with using of 
two-photon methods like to Raman scattering. It is evident that equations 
(\ref{eq17}), (\ref{eq18}) can immediately be used for single transition methods, for 
instance, for IR-absorption. By Raman scattering we have two subsequent 
transitions. Then operator $\hat {\vec {\sigma }}(z)$, which characterizes 
the transition dynamics by Raman scattering process, has to be consisting 
from two components $\hat {\vec {\sigma }}_1 (z)$ and $\hat {\vec {\sigma 
}}_2 (z)$ characterizing both the transitions, taken separately, that is 
$\hat {\vec {\sigma }}(z)=\hat {\vec {\sigma }}_1 (z)+\hat {\vec {\sigma 
}}_2 (z)$. Consequently, the equation for transition dynamics of second 
component, which will determine experimentally observed ESWR-spectrum, is
\begin{gather}
\label{eq19}
\frac{{\partial \hat {\vec {\sigma}} _2 (z)}}{{\partial t}} = \left[ {\hat {\vec {\sigma}} _2 (z) \times \gamma _{_E} \vec E} \right] - \frac{{2a^2 J}}{\hbar}\left[ {\hat {\vec {\sigma}} _2 (z) \times \nabla ^2 \hat {\vec {\sigma}} _2 (z)} \right] \nonumber\\
 - \frac{{2a^2 J}}{\hbar }\left[ {\hat {\vec {\sigma}} _2 (z) \times \nabla ^2 \hat {\vec {\sigma}} _1 (z)} \right].
 \end{gather}
The second and the third items in eq.19 are practically equal to each 
other, since we are dealing with interacting electric dipole moments of the 
same chain, that is, $\nabla ^2\hat {\vec {\sigma }}_1 (z)$ and $\nabla 
^2\hat {\vec {\sigma }}_2 (z)$ have almost equal values. Then we obtain, that 
the value of splitting constant by Raman scattering detection of ESWR in the 
same sample is almost double in comparison with that one by IR detection of 
ESWR. The observation of doubling in the splitting constant by Raman 
ESWR-studies is additional direct argument in ESWR identification.

Therefore, quantum-mechanical analogue of Landau-Lifshitz equation has been derived with clear physical sense of the quantities for both radio- and optical spectroscopy. It has been established that Landau-Lifshitz equation is fundamental physical equation underlying the dynamics of spectroscopic transitions and transitional phenomena. New phenomenon - electrical spin wave resonance and its main properties are predicted. 

The authors are thankfull to Doctor V.Redkov and to Professor, Corresponding Member of National Academy of Sciences of RB L.Tomilchick for the helpful discussions of the part of results.

\bibliography{ref}

\begin{thebibliography}{8}
\expandafter\ifx\csname natexlab\endcsname\relax\def\natexlab#1{#1}\fi
\expandafter\ifx\csname bibnamefont\endcsname\relax
  \def\bibnamefont#1{#1}\fi
\expandafter\ifx\csname bibfnamefont\endcsname\relax
  \def\bibfnamefont#1{#1}\fi
\expandafter\ifx\csname citenamefont\endcsname\relax
  \def\citenamefont#1{#1}\fi
\expandafter\ifx\csname url\endcsname\relax
  \def\url#1{\texttt{#1}}\fi
\expandafter\ifx\csname urlprefix\endcsname\relax\def\urlprefix{URL }\fi
\providecommand{\bibinfo}[2]{#2}
\providecommand{\eprint}[2][]{\url{#2}}

\bibitem[{\citenamefont{Macomber}(1976)}]{Macomber}
\bibinfo{author}{\bibfnamefont{J.~D.} \bibnamefont{Macomber}},
  \emph{\bibinfo{title}{The dynamics of spectroscopic transitions}}
  (\bibinfo{publisher}{John Wiley and Sons}, \bibinfo{address}{New York,
  London, Sydney, Toronto}, \bibinfo{year}{1976}).

\bibitem[{\citenamefont{Scully and Zubairy}(1997)}]{Scully}
\bibinfo{author}{\bibfnamefont{M.~O.} \bibnamefont{Scully}} \bibnamefont{and}
  \bibinfo{author}{\bibfnamefont{M.~S.} \bibnamefont{Zubairy}},
  \emph{\bibinfo{title}{Quantum Optics}} (\bibinfo{publisher}{Cambridge
  University Press}, \bibinfo{address}{Cambridge}, \bibinfo{year}{1997}).

\bibitem[{\citenamefont{Landau and Lifshitz}(1935)}]{Landau_Lifshitz_1935}
\bibinfo{author}{\bibfnamefont{L.~D.} \bibnamefont{Landau}} \bibnamefont{and}
  \bibinfo{author}{\bibfnamefont{E.~M.} \bibnamefont{Lifshitz}},
  \bibinfo{journal}{Physikalische Zeitschrift der Sowjet Union}
  \textbf{\bibinfo{volume}{8}}, \bibinfo{pages}{153} (\bibinfo{year}{1935}).

\bibitem[{\citenamefont{Slichter}(1980)}]{Slichter}
\bibinfo{author}{\bibfnamefont{C.~P.} \bibnamefont{Slichter}},
  \emph{\bibinfo{title}{Principles of Magnetic Resonance}}
  (\bibinfo{publisher}{Sprlnger-Verlag}, \bibinfo{address}{Berlin, Heidelberg,
  New York}, \bibinfo{year}{1980}).

\bibitem[{\citenamefont{Landau and
  Lifshitz}(1981)}]{Landau_Lifshitz_Field_Theory}
\bibinfo{author}{\bibfnamefont{L.~D.} \bibnamefont{Landau}} \bibnamefont{and}
  \bibinfo{author}{\bibfnamefont{E.~M.} \bibnamefont{Lifshitz}},
  \emph{\bibinfo{title}{The Classical Theory of Fields}}
  (\bibinfo{publisher}{Butterworth-Heinemann}, \bibinfo{year}{1981}).

\bibitem[{\citenamefont{Stephani}(2004)}]{Stephani}
\bibinfo{author}{\bibfnamefont{H.}~\bibnamefont{Stephani}},
  \emph{\bibinfo{title}{An Introduction to Special and General Relativity}}
  (\bibinfo{publisher}{Cambridge University Press},
  \bibinfo{address}{Cambridge}, \bibinfo{year}{2004}).

\bibitem[{\citenamefont{Rashevskii}(2003)}]{Rashevskii}
\bibinfo{author}{\bibfnamefont{P.~K.} \bibnamefont{Rashevskii}},
  \emph{\bibinfo{title}{Riemannian Geometry and Tensor Analysis}}
  (\bibinfo{publisher}{Editorial URSS}, \bibinfo{address}{Moscow},
  \bibinfo{year}{2003}).

\bibitem[{\citenamefont{Kittel}(1958)}]{Kittel}
\bibinfo{author}{\bibfnamefont{C.}~\bibnamefont{Kittel}},
  \bibinfo{journal}{Phys.\ Rev.} \textbf{\bibinfo{volume}{110}},
  \bibinfo{pages}{1295} (\bibinfo{year}{1958}).

\end{thebibliography}

\end{document}